\documentclass{jnmp01b}

\usepackage{amsmath}

\numberwithin{equation}{section}

\def\t{{\mathbf t}}                                     %
\def\R{{\mathbf  R}}                                    %
\def\N{{\mathbf  N}}                                    %
\def\Z{{\mathbf  Z}}                                    %
\def\C{{\mathbf  C}}                                    %
\def\A{{\cal A}}                                        %
\def\G{{\cal G}}                                        %
\def\M{{\cal M}}                                        %
\def\H{{\cal H}}                                        %
\def\pa{\partial}                                       %
\def\ker{{{\rm Ker}({\rm  ad\,} \Lambda)}}              %
\def\im{{{\rm  Im}({\rm  ad\,} \Lambda) }}              %
\makeatletter
\DeclareRobustCommand{\primfrac}[1]{%
  \PackageWarning{amsmath}{%
Foreign command \@backslashchar#1; %
\protect\frac\space or \protect\genfrac\space should be used instead%
  }
  \global\@xp\let\csname#1\@xp\endcsname\csname @@#1\endcsname
  \csname#1\endcsname
}
\makeatother

\begin{document}

\renewcommand{\evenhead}{L. Feh\'er and A. G\'abor}
\renewcommand{\oddhead}{Self-Dual Yang-Mills Fields 
in Integrable Hierarchies}

% Title

\thispagestyle{empty}

\begin{flushleft}
\footnotesize \sf
Journal of Nonlinear Mathematical Physics \qquad 2000, V.7, N~4,
\pageref{firstpage}--\pageref{lastpage}.
\hfill {\sc Letter}
\end{flushleft}

\vspace{-5mm}

\copyrightnote{2000}{L. Feh\'er and A. G\'abor}

\Name{A Note on the Appearance of Self-Dual Yang-Mills Fields 
in Integrable Hierarchies}

\label{firstpage}

\Author{L. FEH\'ER and A. G\'ABOR}

\Adress{Department of Theoretical Physics,  University of Szeged, \\
Tisza Lajos krt 84-86, H-6720 Szeged, Hungary\\
E-mail: lfeher@sol.cc.u-szeged.hu}

\Date{Received February 11, 2000; Accepted June 15, 2000}

\begin{abstract}
\noindent
A family of mappings from the solution spaces of certain generalized 
Drinfeld-Sokolov hierarchies to the self-dual Yang-Mills system 
on ${\mathbf R}^{2,2}$  is described.  This provides an extension 
of the well-known relationship between self-dual connections and 
integrable hierarchies of AKNS and Drinfeld-Sokolov type.
\end{abstract}

% The article

\section{Introduction}
                                                                            
The  four-dimensional self-dual Yang-Mills equations 
have several important applications in physics and mathematics \cite{WW}.
For example, many lower dimensional integrable systems can be
obtained as reductions of the self-dual Yang-Mills system, 
which represents a method  to classify integrable systems and 
to apply twistor techniques to study them.
Research in this direction was initiated by Ward \cite{Ward},  
the subsequent investigations are reviewed in \cite{AC,MW}.     
Of particular interest for us is the relationship between the self-dual
Yang-Mills system and  integrable hierarchies of (1+1)-dimensional 
soliton equations.  It has been shown by Mason and collaborators that 
the nonlinear Schr\"odinger and the Korteweg - de Vries 
equations \cite{MSp} as well as the $n$-KdV hierarchies \cite{MSi} 
are reductions of the self-dual 
Yang-Mills system. The generalized AKNS hierarchies \cite{Wil} 
and the modified KdV systems  of Drinfeld-Sokolov \cite{DS} have been 
connected to the self-dual  Yang-Mills system in \cite{McI,Szm, Gu}, 
and super-extensions are discussed in \cite{LaLe}.

It is expected that the set of hierarchies related 
to the self-dual Yang-Mills system is larger than those mentioned 
above, and should include many of the generalized Drinfeld-Sokolov
hierarchies introduced in \cite{GHM}.
To show that this is indeed the case, in this letter 
we exhibit a family of mappings from the solutions 
of a large class of integrable hierarchies to those of the self-dual 
Yang-Mills equations.
We shall be concerned with self-dual Yang-Mills fields on 
$\R^4$ equipped with a pseudo-Euclidean metric of signature (2,2). 
The coordinates on this space are denoted by $y^\mu$ ($\mu=1,2,3,4$)
and the flat metric $\eta_{\mu\nu}$ is given as
\begin{equation}
\eta_{\mu\nu}= \left[ \begin{array}{cccc}
0&0&1&0\\
0&0&0&-1\\
1&0&0&0\\
0&-1&0&0
\end{array}
\right].
\end{equation}
A Yang-Mills field is  represented by a connection  
\begin{equation}
D_\mu = {\partial \over \partial y^\mu} + A_\mu,
\label{I.2}
\end{equation}
where $A_\mu(y)$ varies in some Lie algebra $\G$.
By using $F_{\mu\nu}= [D_\mu, D_\nu]$ and 
the usual Levi-Civita tensor for which $\epsilon_{1234}=1$,
the self-duality equations are defined by 
\begin{equation}
F_{\mu\nu}= {1\over 2} \epsilon_{\mu\nu \alpha \beta} F^{\alpha\beta},
\label{I.3}
\end{equation}
where the indices are raised by the inverse metric $\eta^{\mu\nu}$.
Explicitly, they read as 
\begin{equation}
F_{12}=0,
\qquad
F_{34}=0,
\qquad
F_{13}=F_{24}.
\label{I.4}
\end{equation}
Equivalently, self-duality 
can be characterized by the requirement that the operators 
\begin{equation}
\begin{split}
&P:= D_1 -\lambda D_4 = {\pa \over \partial  y^1} + A_1(y)
- \lambda\left( {\partial \over \partial y^4} + A_4(y)\right) 
 \\
&P':= D_2 -\lambda D_3 =
{\pa \over \partial  y^2} + A_2(y)
- \lambda\left( {\partial \over \partial y^3} + A_3(y)\right) 
\end{split}
\label{I.5}
\end{equation}
satisfy
\begin{equation}
[P,P']=0
\label{I.6}
\end{equation}  
at any value of the spectral parameter $\lambda$.
We shall see below 
that commuting operators of the form $P$, $P'$ above 
naturally arise from a class of integrable hierarchies
which contains, besides the AKNS and Drinfeld-Sokolov hierarchies
associated with the untwisted affine Lie algebras, many other  
special cases of the systems introduced in \cite{GHM}.

\section{Self-dual Yang-Mills from integrable hierarchies}

We next review the basics of the construction of a  
general class of integrable hierarchies.
For a more detailed description, see e.g. \cite{GHM,F}. 
Then we shall specialize the 
input data of the construction in such a way that the solutions of the 
hierarchy give rise to solutions of the self-dual Yang-Mills equation.

The input data of the subsequent construction is a triplet 
$(\A, D,\Lambda)$, where $\A$ is an infinite dimensional Lie algebra,
$D$ is a derivation that defines a ${\bf Z}$-gradation of $\A$, 
and $\Lambda$ is a semisimple element of $\A$ which is homogeneous 
of degree $k>0$ in the gradation $D$.
In more detail, the gradation is given by the eigenspaces $\A^n$
of the derivation $D: \A \rightarrow \A$,
\begin{equation}
\A = \oplus_{n\in \Z} \A^n 
\quad\hbox{with}\quad
\A^n=\{ X\in \A\,\vert\, D(X)=nX\}.
\label{D.5}
\end{equation}
One can restrict the sum to the spectrum of $D$ 
consisting  of the grades for which $\A^n\neq \{ 0\}$,
and we suppose that ${\rm dim}(\A^n)<\infty$.
As a consequence of $D[X,Y]=[DX, Y] + [X, DY]$, one has
$[\A^m, \A^n]\subset \A^{m+n}$.
The semisimplicity of $\Lambda\in \A$ means that it yields
a linear direct sum decomposition as
\begin{equation}
\A=\ker +\im,
\qquad
\ker\cap\im =\{ 0\},
\label{D.3}
\end{equation}
where
\begin{equation}
\ker = \{ X\in \A \,\vert\, [\Lambda, X]=0\},
\quad
\im= \{ X\in \A \,\vert\, X=[\Lambda, Y] \ \hbox{for}\ Y\in \A \}.
\label{D.4}
\end{equation}
The assumption that $D(\Lambda)= k \Lambda$ 
ensures that $\ker$ is a graded Lie subalgebra of $\A$
and $\im$ is a graded vector space.

Given the above data,
one can define an infinite set of commuting  evolutional vector 
fields on the `phase space' 
\begin{equation}
\M_{\Lambda, D}:=\{ L=\pa_x + j(x) + \Lambda\,\vert\, j(x) \in \A^{<k}\}.
\label{MLaD}
\end{equation}
Here $\A^{<k}=\oplus_{n<k} \A^n$ and analogous 
notations will be used without further explanation. 
The dynamical variables are the infinitely many independent component
fields of $j(x)$, which are smooth  
(real or complex valued depending on whether
$\A$ is a real or a complex Lie algebra)
functions of the one-dimensional `space variable' $x$. 
The evolutional vector fields are associated with 
the non-negatively graded part $\H_\Lambda^+$ of the centre of $\ker$:
\begin{equation}
\H_\Lambda^+ = \{ c\in \ker\,\vert\, [c,X ]= 0 
\quad \forall X\in \ker\,\}\cap \A^{\geq 0}.
\end{equation}
Concretely, for any $c\in \H_\Lambda^+$ one obtains   
a vector field $V_c$ on $\M_{\Lambda,D}$ as follows.
First, for any $L\in \M_{\Lambda, D}$ 
one considers the Drinfeld-Sokolov dressing transformation:
\begin{equation}
L= (\pa_x + j(x) + \Lambda)  \mapsto e^{{\rm ad} F}(L)=
(\pa_x + h(x) + \Lambda),
\label{D.8}
\end{equation}
where  $F(x)$ and $h(x)$ are supposed to be formal series 
\begin{equation}
F(x)\in \im^{<0},
\qquad
h(x)\in \left(\ker\right)^{<k}.
\label{D.9}
\end{equation}
It is well-known \cite{DS,GHM} that there exists a unique 
solution $F(x)=F(j(x))$, $h(x)=h(j(x))$ of (\ref{D.8}) for any $L$,
and the components of $F$ and $h$ are finite polynomials 
in the components of $j(x)$ and their $x$-derivatives.
We here use  
\begin{equation}
e^{{\rm ad} F}(L)=\sum_{n=0}^\infty {1\over n!} ({\rm ad} F)^n(L)
\quad\hbox{with}\quad
({\rm ad} F)(L)=-\pa_x F + [F,j+\Lambda],
\label{D.11}
\end{equation}
or, equivalently, $e^{{\rm ad} F}(L)$ may be thought of 
as $e^F L e^{-F}$.
The vector field $V_c$ is defined by its action on 
the `coordinate' $j$ on $\M_{\Lambda,D}$ as follows:
\begin{equation}
V_c j(x) := 
-\pa_x B(c, j(x))^{\geq 0} + [B(c,j(x))^{\geq 0}, j(x) + \Lambda],
\label{Vc}
\end{equation}
where  
\begin{equation}
B(c,j):=e^{-{\rm ad} F(j) } (c)
\qquad \forall c\in \H^{+}_\Lambda,
\label{Bcj}
\end{equation}
and  $B(c,j)= B(c,j)^{\geq 0} + B(c,j)^{<0}$ according 
to the gradation $D$. 
Since $[ B(c,j)^{\geq 0}, L] =- [B(c,j)^{<0},L]$, $V_c$ 
is a well-defined vector field on $\M_{\Lambda,D}$, that is, 
$V_c(j(x))\in \A^{<k}$.
Since $\H_\Lambda^+$ is an Abelian 
Lie algebra, the vector fields $V_c$ commute among themselves,
\begin{equation}
[V_c, V_{c'}] (j(x)) = 0 
\qquad
\forall c, c'\in \H_{\Lambda}^{+}.
\label{Vcommute}
\end{equation}
If we introduce a time variable $t^c$ for every basis element
$c\in \H_\Lambda^{+}$, and denote the collection of these
variables by $\t$, 
 then we can write the  
evolution equations defined by the $V_c$,
\begin{equation}
{\pa \over \pa t^c} j(x,\t)= 
-\pa_x B(c, j(x,\t))^{\geq 0} + 
[B(c,j(x,\t))^{\geq 0}, j(x,\t) + \Lambda],
\label{tceq}
\end{equation}
in the zero curvature form 
\begin{equation}
[{\cal L}_c, L]=0
\qquad\hbox{with}\qquad 
L=\pa_x + j(x,\t) +\Lambda,
\qquad
{\cal L}_c = {\pa \over \partial t^c} - B(c,j(x,\t))^{\geq 0}.
\label{LcL}
\end{equation}
As a result of the compatibility (\ref{Vcommute}) of these 
evolution equations, one then also has
\begin{equation}
[{\cal L}_c, {\cal L}_{c'} ]=0
\qquad
\forall c, c'\in \H_{\Lambda}^{+}.
\end{equation}

After this preparation, we now show that under certain choices 
of the data the above zero curvature equations imply the self-duality 
equations described in the introduction.
Since we wish to obtain commuting operators of the form in (\ref{I.5}),
we take $\A$ to be a loop algebra, 
\begin{equation}
\A= \G \otimes \C((\lambda^{-1}))= 
\{X=\sum_{i\leq N_X} X_i \lambda^i\,\vert\,  
X_i \in \G,\ N_X\in \Z\,\},
\label{loopalg}
\end{equation}
where $\G$ is a finite dimensional Lie algebra.
We assume that the gradation $D$ of $\A$ is given by 
\begin{equation}
D= M \lambda {d\over d\lambda} + {\rm ad} H,
\label{gradD}
\end{equation}
where $H$ is an ad-diagonalizable element of $\G$ with integer eigenvalues
whose absolute values are  {\em strictly less} than $M\in \N$.
Thus we have a corresponding gradation of $\G$,
\begin{equation}
\G = \oplus_{n} \G^n 
\quad\hbox{where}\quad 
\vert n \vert  < M.
\label{gradG}
\end{equation}
As before, $\Lambda\in \A$ is a semisimple element 
of $D$-grade $k>0$.
Under these assumptions, we can introduce a basis 
$c^{a,l}$ of $\H_\Lambda^+$ 
given by 
\begin{equation}
c^{a, l} := \lambda^l c^{a} 
\quad\hbox{with}\quad c^a\in \H_\Lambda^+ \cap \A^{<M},
\quad\hbox{and}\quad
l=0,1,2,\ldots,
\end{equation}
where $c^a=c^{a,0}$ is a basis of $\H_\Lambda^+ \cap \A^{<M}$.
We shall denote the time variables $t^c$ and the Lax operators ${\cal L}_c$ 
corresponding to these base elements 
respectively as $t^{a,l}$ and ${\cal L}_{a,l}$.
The general construction now implies that 
\begin{equation}
B(c^{a,l},j) = \lambda^l B(c^a,j),
\end{equation}
and it follows from our assumption on the gradation 
of $\G$ in (\ref{gradG}) that
$B(c^a,j)$ expands as  
\begin{equation}
B(c^a,j)= \sum_{i\leq 1} B_i(c^a,j) \lambda^i,
\qquad
B_i(c^a,j)\in \G.
\end{equation} 
By using these expansions, the Lax operators defining 
the hierarchy can be written as 
\begin{equation}
{\cal L}_{a,l} = {\pa \over \pa t^{a,l}} - B_{-l}(c^a, j(x,\t))^{\geq 0} - 
\sum_{i=1-l}^{1} \lambda^{i+l} B_i(c^a, j(x,\t))
\qquad\forall l\geq 0.
\label{Lal}
\end{equation}
In particular, these operators contain only non-negative
powers of $\lambda$, due to our assumption on the 
eigenvalues of ${\rm ad} H$
in (\ref{gradG}).
We then notice that the operators 
$P_{a,l}:={\cal L}_{a,l+1} - \lambda {\cal L}_{a,l}$ have {\em the same form
as the Lax operators of the self-dual Yang-Mills 
equations in} (\ref{I.5}).
Indeed, we obtain 
\begin{equation}
P_{a,l}={\pa \over \pa t^{a,l+1}} - B_{-l-1}(c^a, j(x,\t))^{\geq 0}
-\lambda\left(
{\pa \over \pa t^{a,l}} + B_{-l}(c^a, j(x,\t))^{<0}\right)
\qquad\forall  l\geq 0,
\label{Pal}
\end{equation} 
where 
$B_i(c^a, j(x,\t))^{\geq 0}$ (resp. $B_i(c^a, j(x,\t))^{<0}$) is 
the part of $B_i(c^a, j(x,\t))\in \G$ 
defined by the decomposition $\G= \G^{\geq 0} + \G^{<0}$
associated with (\ref{gradG}). 
Moreover, 
it is convenient to introduce `cyclic coordinates' $t^{a, -1}$, 
on which nothing depends,
since by using  them we may write 
\begin{equation}
P_{a,-1} :={\cal L}_{a,0}-\lambda {\pa \over \pa t^{a,{-1}}} =
{\pa \over \pa t^{a,0}} - B_{0}(c^a, j(x,\t))^{\geq 0}
-\lambda\left(
{\pa \over \pa t^{a,-1}} + B_{1}(c^a, j(x,\t))\right),
\label{Pa0}
\end{equation}
which formally follows the same pattern as the other $P_{a,l}$.
Note that $B_1(c^a,j(x,\t))\in \G^{<0}$. 

Let us now choose pairs of indices $a,l$ and $a',l'$ in such a way 
that the set 
\begin{equation}
y^1:=t^{a,l+1},\qquad
y^4:=t^{a,l},\qquad  
y^2:=t^{a', l'+1},\qquad
y^3:=t^{a',l'}
\label{ty}
\end{equation}
consists of 4 different coordinates.
The main point of this paper is the observation that {\em any 
solution $j(x,\t)$ of the above constructed integrable hierarchy
gives rise to self-dual Yang-Mills fields by the mapping}
\begin{equation}
 P_{a,l}\mapsto P,
\qquad 
P_{a', l'}\mapsto P'.
\label{PPprime}
\end{equation}  
To obtain self-dual Yang-Mills fields by this correspondence,
it is enough that the evolution equations of the
hierarchy be satisfied for the finite subset of the variables given by $x$ 
and the $t$'s that occur in (\ref{ty}).
On the other hand, if $j(x,\t)$ satisfies the  
hierarchy in all (or many) time variables, then it gives rise
to (infinitely) many self-dual Yang-Mills fields, too, since the
identification of the $y^\mu$'s in terms of the $t$'s is then possible 
in several different ways.
Note that if $P_{a,-1}$ is taken as one of the Lax operators,
then the corresponding Yang-Mills fields are invariant
under the translations of the cyclic coordinate
$t^{a,-1}$, and thus the component $B_1(c^a, j(x,\t))$
associated with this coordinate in (\ref{Pa0}) 
can be regarded as a `Higgs field' in the adjoint
representation of $\G$. 
In these cases the self-duality equation (\ref{I.6}) 
becomes an analogue 
of the Bogomolny equation (see e.g.~\cite{MW}).

The self-dual Yang-Mills fields that appear in the above 
are special since they vary 
in the subalgebras $\G^{\geq 0}$ and $\G^{<0}$ of $\G$.
If $D$ is a multiple of the homogeneous 
gradation, $D=M\lambda {d\over d\lambda}$, then $\G^{\geq 0}=\G$ and
$\G^{<0}=\{0\}$, and in this case another special 
feature  is that the Yang-Mills
potentials that multiply $\lambda$ in $P_{a,l}$ vanish.
The Yang-Mills potentials that arise from 
(\ref{PPprime}) are differential polynomials
in the infinitely many components of the field 
$j(x,\t)\in \A^{<k}$ contained 
in the basic Lax operator $L$ of the hierarchy. 
After reducing the degrees of freedom to finitely many fields,
in certain cases $j$ itself can be regarded as a $\G$-valued 
Yang-Mills potential.
These reductions are discussed briefly in the next section.

\section{Reductions, gauge transformations and examples}
\resetfootnoterule

It is not difficult to check that the vector fields $V_c$ (\ref{Vc})
are always tangential to the submanifold 
$\Theta_{\Lambda, D}\subset \M_{\Lambda,D}$ defined by
\begin{equation}
\Theta_{\Lambda, D}
:=\{ L=\pa_x + j(x) + \Lambda\,\vert\, j(x) \in \A^{<k}\cap \A^{\geq 0}\}.
\label{Theta}
\end{equation}
The restriction of the evolution equations (\ref{tceq}) 
to this phase space, where $j$ has only finitely many components, 
is called a modified  KdV type hierarchy \cite{GHM}. 
Among the modified KdV type hierarchies that can be mapped to the 
self-dual Yang-Mills system,
let us now concentrate on those for which the condition 
\begin{equation}
D(\Lambda) = k\Lambda
\qquad
0<k\leq M
\end{equation}
holds besides eqs.~(\ref{loopalg}), (\ref{gradD}), (\ref{gradG}).
Then, introducing also another cyclic coordinate $z$, 
we can represent the basic Lax operator $L$ as
\begin{equation}
L= {\pa \over \pa x} + (j_0(x,\t) + \Lambda_0) - \lambda
\left({\pa \over \pa z} - j_1(x,\t) - \Lambda_1\right),
\end{equation} 
since we have the decompositions 
$\Lambda=(\Lambda_0 + \Lambda_1 \lambda)$
and $j= (j_0 + j_1 \lambda)$ with some $\Lambda_i, j_i \in \G$.
Notice that this operator now has the same form as those in (\ref{I.5}).
Therefore it follows from (\ref{LcL}) that 
we obtain self-dual Yang-Mills fields from the 
solutions of the hierarchy on $\Theta_{\Lambda,D}$ not only 
by the correspondence 
in (\ref{PPprime}), but also via the mapping
\begin{equation}
P_{a,l}\mapsto P,
\qquad
 L\mapsto P',
\label{PL}
\end{equation}
whereby $(j_0 + \Lambda_0)$ and $-(j_1 + \Lambda_1)$ 
become components of the Yang-Mills potential.
More precisely, $(j_1 + \Lambda_1)$ can be regarded as a Higgs field 
since nothing depends on the associated space-time coordinate $y^3=z$.
Examples of this kind are the generalized AKNS systems
for which  $k=M=1$ with $D=\lambda {d\over d\lambda}$, 
$\Lambda=\lambda \Lambda_1$.
The mapping of the solutions of the hierarchy to self-dual connections 
has in this case been described previously in \cite{McI}.
Another set of examples is furnished by 
the generalized modified KdV systems of Drinfeld-Sokolov 
\cite{DS} associated with the untwisted affine Lie algebras.
For these systems $D$ is the so called principal gradation 
and $\Lambda$ has grade  $k=1<M$,
where $M$ is the dual Coxeter number of a simple Lie algebra $\G$.

A different class of reductions of the systems on $\M_{\Lambda,D}$ 
are the KdV type hierarchies \cite{DS,GHM}. 
Their construction uses,  
in addition to $\A$ in (\ref{loopalg}), $D$ in (\ref{gradD}) and 
 $\Lambda\in\A$  of $D$-grade $k>0$,
an auxiliary integral gradation of $\A$.
Let this be provided by the eigenvalues of the derivation 
\begin{equation}
d = m \lambda {d\over d\lambda} + {\rm ad} h
\qquad\quad (m\in \N,\, h\in \G).
\end{equation}
The two gradations are compatible, $[D,d]=0$, and
must satisfy  
\begin{equation}
\A^{D>0}\subset \A^{d\geq 0},
\qquad
\A^{D=0}\subset \A^{d=0},
\qquad
\A^{D<0}\subset \A^{d\leq 0}.
\label{Ddcond}
\end{equation}
Furthermore, one assumes that 
\begin{equation}
\Gamma := \A^{D<0} \cap \A^{d=0}
\end{equation}
is a non-trivial subalgebra of $\A$ for which 
\begin{equation}
\Gamma \cap \ker = \{ 0\}.
\label{nondeg}
\end{equation}
For our purpose, we also require that $\vert {\rm ad} h \vert <m$,
which means that the ${\rm ad} h$-grades in 
$\G$ are strictly smaller 
than $m$ (analogously to the property of $D$ in (\ref{gradG})).
As a consequence, 
we have $\Gamma \subset \G$, where $\G$ is naturally 
embedded in $\A$ as the set of $\lambda$-independent elements.

To associate a KdV type hierarchy with the
data $(\A, \Lambda, D,d)$, one first restricts the 
hierarchy on $\M_{\Lambda,D}$ to the submanifold
\begin{equation}
 \M_{\Lambda,D,d}:=
\{ L=\pa_x + j(x) + 
\Lambda\,\vert\, j(x) \in \A^{D<k}\cap \A^{d\geq 0}\}
\subset \M_{\Lambda,D}.
\label{MDd}
\end{equation}
The above conditions guarantee that the restriction is consistent.
That is the vector fields $V_c$ (\ref{Vc}),  
still defined by means of the $D$-gradation,
are in our case tangential to  $\M_{\Lambda,D,d}$.
The next step is to consider the transformations
\begin{equation}
L=(\pa_x + j(x) + \Lambda)  \mapsto L^\gamma= 
e^{\gamma(x)} L e^{-\gamma(x)}=(\pa_x + j^\gamma(x) +\Lambda)  
\qquad
\forall \gamma(x) \in \Gamma.
\label{Lgamma}
\end{equation}
Explicitly, 
\begin{equation}
j^\gamma(x) = e^{\gamma(x)} (j(x) +\Lambda) e^{-\gamma(x)}
-\Lambda - (\pa_x e^{\gamma(x)})e^{-\gamma(x)}.
\label{jgamma}
\end{equation}
These transformations define an action of the group 
${\cal N}:= \{ e^\gamma \,\vert\, \gamma(x) \in \Gamma\}$
on $\M_{\Lambda,D,d}$,
and one sees from (\ref{nondeg}) that this is a free action,
i.e., only the unit element of ${\cal N}$  has  fixed points. 
Then ${\cal N}$ is regarded as a group of {\em gauge transformations} 
for the hierarchy on  $\M_{\Lambda,D,d}$, which is 
possible because the  
vector field $V_c$ (\ref{Vc}) has a well-defined projection 
on the factor space $\M_{\Lambda,D,d}/{\cal N}$.
This follows from the  relations 
\begin{equation}
B(c, j^\gamma) = e^\gamma B(c,j) e^{-\gamma}
\quad\hbox{and}\quad
\left(e^\gamma B(c,j)^{D\geq 0} e^{-\gamma} - 
B(c,j^\gamma)^{D\geq 0}\right)\in \Gamma, 
\label{equiv}
\end{equation}
which are consequences of the definition of $B(c,j)$ (\ref{Bcj})  
and eqs.~(\ref{Ddcond}), (\ref{nondeg}). 
By definition,
the KdV type hierarchy is the hierarchy 
on $\M_{\Lambda, D,d}$ modulo the 
gauge group ${\cal N}$.

As this is known to hold in special cases \cite{MSp, MSi},
it is natural to ask whether 
`KdV gauge equivalence' always implies 
gauge equivalence of the corresponding 
self-dual Yang-Mills fields.
To answer this question,  
consider a time dependent gauge transformation 
of a solution 
of the hierarchy on $\M_{\Lambda,D,d}$.
The solution represented by $L(\t)=(\pa_x + j(x,\t) +\Lambda)$  is sent to 
 $L^\gamma(\t)=(\pa_x + j^\gamma(x,\t) +\Lambda)$,
where we apply (\ref{Lgamma}) with a function $\gamma(x,\t)\in \Gamma$.
The definition of the hierarchy implies that $j^\gamma(x,\t)$ 
will be another solution if and only if 
$\gamma(x,\t)$ solves the following evolution equations: 
\begin{equation}
{\pa e^{\gamma(x,\t)} \over \pa t^c} e^{-\gamma(x,\t)}
= B(c, j^\gamma(x,\t))^{D\geq 0} - e^{\gamma(x,\t)} 
B(c, j(x,\t))^{D\geq 0}\,
 e^{-\gamma(x,\t)}.
\label{gammat}
\end{equation}
By using (\ref{equiv}) and (\ref{Vcommute}), one can check 
that (\ref{gammat}) yields well-defined, compatible evolution equations
on $\gamma(x,\t)$. 
In conclusion, the solutions of the hierarchy on $\M_{\Lambda,D,d}$
which are gauge equivalent to $L(\t)$ have the form 
$L^\gamma(\t)$ where $\gamma(x,\t)$ solves (\ref{gammat}) 
for some initial value $\gamma(x)$ at $\t=0$.
One easily sees from this that the Lax operators 
${\cal L}_c$ defined in (\ref{LcL})  
behave in the following manner with respect to 
the gauge transformations of the solutions:
\begin{gather}
{\cal L}_c=\left( {\pa \over \pa t^c} - B(c, j(x,\t))^{D\geq 0}\right)
\mapsto \notag \\
\qquad
{\cal L}_c^{\gamma}:= 
\left( {\pa \over \pa t^c} - B(c, j^\gamma(x,\t))^{D\geq 0}\right)
= e^{\gamma(x,\t)} {\cal L}_c e^{-\gamma(x,\t)}.
\label{KdVgauge}
\end{gather}
This formula permits us to  show that  
the self-dual Yang-Mills fields corresponding,  
either\footnote{Due to our assumptions on the eigenvalues 
of ${\rm ad} H$
and ${\rm ad}\, h$, the mapping (\ref{PL}) can be used 
if $k\leq M$ is valid for the KdV type system.}
by  (\ref{PPprime}) or by (\ref{PL}),  
to gauge equivalent solutions of the 
hierarchy on $\M_{\Lambda,D,d}$ 
{\em are always gauge equivalent in the usual Yang-Mills sense}.
The equivalence of two $\G$-valued Yang-Mills connections 
$D_\mu$ and $\tilde D_\mu$ 
requires the existence of a function
$g(y)$ in the group generated by $\G$ for which 
$\tilde D_\mu = g(y) D_\mu g^{-1}(y)$.
In the self-dual case, this amounts to  
$\tilde P(y)= g(y) P(y) g^{-1}(y)$
and to $\tilde P'(y)= g(y) P'(y) g^{-1}(y)$ for the operators
in (\ref{I.5}).
Now it is obvious  from (\ref{KdVgauge}) and (\ref{Pal}), (\ref{Pa0})
that if $D_\mu$ belongs
to $j(x,\t)$ and $\tilde D_\mu$ to $j^\gamma(x,\t)$ then 
they are gauge related by 
$g(y)=e^{\gamma(x,\t)}$, where the arguments of 
$\gamma(x,\t)$ different from the $y^\mu$ are fixed arbitrarily. 
As mentioned above, $\Gamma \subset \G$, and thus the proof is complete.

The conditions used in the above are satisfied, for example, 
for the generalized KdV hierarchies associated 
with the regular conjugacy classes of the Weyl groups of the simple 
Lie algebras \cite{DF}.
These are special cases of the systems introduced in \cite{GHM},
and contain the untwisted Drinfeld-Sokolov hierarchies, 
which belong to the Coxeter conjugacy class.   
Most of them can be described alternatively as constrained 
(matrix) KP hierarchies.
As one of the simplest examples, let us mention the 
much studied constrained KP
hierarchies that have scalar Lax operators of the form
\begin{equation}
L_{\rm cKP}= \pa^n + u_1\pa^{n-1} +\cdots + u_n + v_+ (\pa + w)^{-1} v_-
\end{equation}
for some $n\geq 1$.
These systems are associated with $gl_{n+1}$ and the elements of the 
corresponding conjugacy class
of the Weyl group $S_{n+1}$ decompose as the product of 
a $n$-cycle and a $1$-cycle.
For detailed description, see e.g.~\cite{FM} and references therein. 
It would be interesting to further explore the 
nature of the relationship between these hierarchies 
and the self-dual Yang-Mills equations.
For instance, following the spirit of \cite{MW,MSp,MSi},  
it should be possible to view at least some 
of them as reductions of a self-dual Yang-Mills hierarchy,
and to apply twistor techniques to describe their $\tau$-functions.   

Finally, we wish to remark that
there exist certain constrained modified KP 
(also called non-standard  $n$-KdV)  hierarchies that fit into 
an extended version of the Drinfeld-Sokolov approach for which our above 
assumptions do not hold \cite{DFG}. 
These systems, and also the Drinfeld-Sokolov hierarchies  based 
on the twisted affine Lie  algebras, 
cannot be mapped to the self-dual Yang-Mills
system by the method presented here.
In fact, in these cases the analogues of
$P_{a,l}={\cal L}_{a,l+1} - \lambda {\cal L}_{a,l}$ in (\ref{Pal}) would contain 
both positive and negative,  or higher than first
powers of the spectral parameter, which do not appear in (\ref{I.5}).

\subsection*{Acknowledgements}
This work has been supported in part by the Hungarian Ministry of Education
under  FKFP 0596/1999 and by the National Science Fund (OTKA) under T025120.

\label{lastpage}

\end{document}